\documentclass[11pt,onecolumn]{article}
\usepackage[margin=2.2cm]{geometry}
\makeatletter
\@ifclassloaded{revtex4-1}{}{ \usepackage[square,numbers]{natbib}\bibliographystyle{ieeetr} }
\makeatother % alpha,apsrev4-1,ieeetr
\usepackage[usenames,dvipsnames]{xcolor} % \usepackage[utf8]{inputenc}
\usepackage{hyperref,lipsum,enumitem,graphicx,csquotes,tikz,pgfplots,makecell,pgffor,catchfilebetweentags,mdframed} % makecell for table; pgffor for foreach in preamble
\graphicspath{{/home/wei/Dropbox/tex/fig/}} % \graphicspath{{fig/}}
\usepackage[smalltableaux]{ytableau}
\usepackage{bbm,bm,mathrsfs,amsmath,amsthm,amssymb,mathtools,dsfont} % math
\usepackage[titletoc,title]{appendix}
\usetikzlibrary{cd,arrows,decorations.pathmorphing,backgrounds,positioning,fit,matrix,plotmarks} %positioning sharelatex

\newcommand{\nc}{\newcommand}               \nc{\rnc}{\renewcommand}

    \nc{\bt}{\beta}          \nc{\gm}{\gamma}  \nc{\Gm}{\Gamma} \nc{\dt}{\delta}
\nc{\Dt}{\Delta}   \nc{\kp}{\kappa}         \nc{\sg}{\sigma}  \nc{\Sg}{\Sigma} \nc{\tht}{\theta}
\nc{\Tht}{\Theta}  \nc{\ld}{\lambda}        \nc{\Ld}{\Lambda} \nc{\om}{\omega} \nc{\Om}{\Omega}
\nc{\phv}{\varphi} \nc{\epsl}{\varepsilon}  \nc{\thv}{\vartheta} 
\nc{\s}{\mathsf}                            \nc{\domathsf}[1]{\s{#1}}    % math sans serif letters \As \Bs
\nc{\xs}{\textsf}                           \nc{\dotextsf}[1]{\xs{#1}}   % text sans serif letters \As \Bs
\nc{\x}{\textnormal}                        \nc{\dotextnormal}[1]{\x{#1}}% text font in math \Ax \Bx
\nc{\n}{\operatorname}                      \nc{\dooperatorname}[1]{\n{#1}}
\nc{\domathcal}[1]{\mathcal{#1}}                                         % Calligraphic letters \Ac \Bc
\nc{\domathfrak}[1]{\mathfrak{#1}}                                       % Fraktur letters \gf \hf
\rnc{\b}[1]{\mathbf{#1}}                    \nc{\xb}[1]{\textbf{#1}}     % Upright bold; or \bm{}
\def\capitalletters{A,B,C,D,E,F,G,H,I,J,K,L,M,N,O,P,Q,R,S,T,U,V,W,X,Y,Z,f}
\def\inmathsf{SL,SO,Sp,SU,BQP,GL,Gr,Seg}
\def\intextnormalx{C,D,E,F,H,I,L,P,Q,R,U}
\def\intextnormal{conv,CP,diag,End,Herm,Hom,HS,im,Pd,Pos,Var,poly,spec,LOCC,Sep,PPT,%
                  rank,rk,supp,sgn,vol,Wg,id}
\def\inoperatorname{Tr,tr}
\def\inmathfrak{g,h,m,u,E,I,J,l,L,M,U}
\def\inmathfraknof{gl,sl,so,sp,su}
\foreach \i in \capitalletters { \expandafter\xdef\csname \i s\endcsname{ \noexpand\domathsf{\i} } }
\foreach \i in \capitalletters { \expandafter\xdef\csname \i xs\endcsname{ \noexpand\dotextsf{\i} } }
\foreach \i in \capitalletters { \expandafter\xdef\csname \i c\endcsname{ \noexpand\domathcal{\i} } }
\foreach \i in \capitalletters { \expandafter\xdef\csname c\i \endcsname{ \noexpand\domathcal{\i} } } %%x
\foreach \i in \inmathfrak { \expandafter\xdef\csname \i f\endcsname{ \noexpand\domathfrak{\i} } }
\foreach \i in \inmathfraknof { \expandafter\xdef\csname \i \endcsname{ \noexpand\domathfrak{\i} } }
\foreach \i in \inmathsf { \expandafter\xdef\csname \i \endcsname{ \noexpand\domathsf{\i} } }
\foreach \i in \intextnormal { \expandafter\xdef\csname \i \endcsname{ \noexpand\dotextnormal{\i} } }
\foreach \i in \intextnormalx { \expandafter\xdef\csname \i x\endcsname{ \noexpand\dotextnormal{\i} } }
\foreach \i in \inoperatorname { \expandafter\xdef\csname \i \endcsname{ \noexpand\dooperatorname{\i} } }
\nc{\1}{\mathds{1}}  \nc{\Bb}{\mathbb{B}}   \nc{\Cb}{\mathbb{C}} \nc{\Fb}{\mathbb{F}} \nc{\Hb}{\mathbb{H}}
\nc{\Nb}{\mathbb{N}} \nc{\Rb}{\mathbb{R}}   \nc{\Sb}{\mathbb{S}} \nc{\Zb}{\mathbb{Z}}
\rnc{\P}{\mathbb{P}} \nc{\E}{\mathbb{E}}    \nc{\Eb}{\mathop{{}\mathbb{E}}}
\nc{\RR}{\mathbf{R}} \nc{\rr}{\mathbf{r}}   \nc{\cc}{\mathbf{c}} \nc{\sss}{\mathbf{s}}
\nc{\Bscr}{\mathscr{B}} \nc{\Dscr}{\mathscr{D}} \nc{\Uscr}{\mathscr{U}}

\nc{\ot}{\otimes}                           \nc{\ox}{\ot} %%x
\nc{\T}{^{\s{T}}}                           \nc{\PT}[1]{^{\s{T}_{\!{#1}}}}
\nc{\Ot}{\bigotimes}                        \nc{\dg}{^{\dagger}}
\nc{\Oplus}{\bigoplus}                      \nc{\f}[2]{\frac{#1}{#2}}
\nc{\wtd}{\widetilde}  \nc{\wht}{\widehat}  \nc{\ovl}{\overline} 
\nc{\bra}[1]{\langle{#1}|}                  \nc{\brak}[1]{\langle{#1}\rangle}
\nc{\ket}[1]{|{#1}\rangle}                  \nc{\ketb}[2]{|{#1}\rangle\!\langle{#2}|}
\nc{\ps}[1]{\left(#1\right)}                \nc{\pv}[1]{\left|#1\right|}
\nc{\pss}[1]{\left[#1\right]}               \nc{\pvv}[1]{\left\|#1\right\|}
\nc{\psss}[1]{\left\{#1\right\}}            % \nc{\jquote}[1]{$\lceil$#1$\rfloor$}
\nc{\fl}[1]{\left\lfloor#1\right\rfloor}    \nc{\xto}[1]{\xrightarrow{#1}}
\nc{\spn}{\x{span}}   \nc{\vc}{\n{vec}}     \nc{\vecc}{\overrightarrow}
\nc{\ex}{\x{e}^}   \nc{\dd}{\,\x{d}}        \nc{\ee}{\x{e}}   \nc{\ii}{\x{i}} 
\nc{\2}{\{0,1\}}							\nc{\lims}{\varlimsup} \nc{\limi}{\varliminf}

\nc{\Frm}[1]{\begin{mdframed}[backgroundcolor=blue!10,hidealllines=true]#1\end{mdframed}}
\nc{\Frmd}[1]{\begin{mdframed}[backgroundcolor=blue!10,linecolor=blue,linewidth=1,innerleftmargin=2,innerrightmargin=2,innertopmargin=0,innerbottommargin=0]#1\end{mdframed}}

\newtheorem{thm}{Theorem}                   \newtheorem{prop}[thm]{Proposition}
\newtheorem{defi}[thm]{Definition}          \newtheorem{coro}[thm]{Corollary}
\newtheorem{lem}[thm]{Lemma}                

\nc{\Thm}[1]{\begin{thm}#1\end{thm}}
\nc{\Def}[1]{\begin{defi}#1\end{defi}}
\nc{\Lem}[1]{\begin{lem}#1\end{lem}}
\nc{\Prop}[1]{\begin{prop}#1\end{prop}}
\nc{\Coro}[1]{\begin{coro}#1\end{coro}}
\nc{\Eq}[1]{\begin{equation}#1\end{equation}}
\nc{\Eqn}[1]{\begin{equation*}#1\end{equation*}}
\nc{\Al}[1]{\begin{align}#1\end{align}}
\nc{\Aln}[1]{\begin{align*}#1\end{align*}}
\nc{\Als}[1]{\begin{align}\begin{split}#1\end{split}\end{align}}
\nc{\Item}[1]{\begin{itemize}#1\end{itemize}}
\nc{\Enum}[1]{\begin{enumerate}#1\end{enumerate}} % add [nolistsep] before #1 to nolistsep
\nc{\Desc}[1]{\begin{description}#1\end{description}}

\nc{\arr}[1]{\begin{array}#1\end{array}}
\nc{\mm}[2]{\ps{\arr{{c}{#1}\\{#2}}}}
\nc{\mms}[2]{\pss{\arr{{c}{#1}\\{#2}}}}
\nc{\mmm}[3]{\ps{\arr{{c}{#1}\\{#2}\\{#3}}}}
\nc{\mmmm}[4]{\ps{\arr{{c}{#1}\\{#2}\\{#3}\\{#4}}}}
\nc{\mmnn}[4]{\ps{\arr{{cc}{#1}&{#2}\\{#3}&{#4}}}}
\nc{\mmmnn}[6]{\ps{\arr{{cc}{#1}&{#2}\\{#3}&{#4}\\{#5}&{#6}}}}
\nc{\mmnnn}[6]{\ps{\arr{{ccc}{#1}&{#2}&{#3}\\{#4}&{#5}&{#6}}}}
\nc{\mmmnnn}[9]{\ps{\arr{{ccc}{#1}&{#2}&{#3}\\{#4}&{#5}&{#6}\\{#7}&{#8}&{#9}}}}
\nc{\ydiag}[1]{\ytableausetup{notabloids}\ydiagram{#1}}
\nc{\ytab}[1]{\ytableausetup{notabloids}\ytableaushort{#1}}
\nc{\ytabb}[1]{\ytableausetup{tabloids}\ytableaushort{#1}}

\definecolor{mygray}{gray}{0.5}
\nc{\gray}[1]{{\color{mygray}#1}}
\nc{\blue}[1]{{\color{blue}#1}}
\nc{\emm}[1]{{\color{blue}\em #1}}
\nc{\red}[1]{{\color{red}#1}}
\definecolor{darkblue}{rgb}{0.1,0.15,0.7}
\usepackage{hyperref}
\hypersetup{colorlinks=true,citecolor=darkblue,linkcolor=darkblue,filecolor=darkblue,urlcolor=darkblue,breaklinks=true}

\newcommand{\proj}[1]{| #1\rangle\!\langle #1 |}

\usepackage[affil-it]{authblk}
\title{Converse bounds for classical communication over quantum networks}
\author[]{Wei Xie\thanks{\href{mailto:xievvvei@gmail.com}{xievvvei@gmail.com}} }
\author[]{Xin Wang\thanks{\href{mailto:xin.wang-8@student.uts.edu.au}{xin.wang-8@student.uts.edu.au}} }
\author[]{Runyao Duan\thanks{\href{mailto:runyao.duan@uts.edu.au}{runyao.duan@uts.edu.au}}}
\affil[]{Centre for Quantum Software and Information,\\Faculty of Engineering and Information Technology, University of Technology Sydney, NSW 2007, Australia}

\begin{document}
% \author{Wei Xie} \email{xievvvei@gmail.com}
% \author{Xin Wang} \email{xin.wang-8@student.uts.edu.au}
% \affiliation{Centre for Quantum Software and Information, Faculty of Engineering and Information Technology, University of Technology Sydney, NSW 2007, Australia}
% \title{Strong converse bounds for classical communication over quantum networks}

\date{\today}\maketitle
\begin{abstract}
%%%%%%%%%%%%%%%%%%%%%%%%%%%%%to be deleted%%%%%%%%
%\XW{Contributions of this paper:
%\begin{itemize} 
%\item Generalize NSPPT for communication from point-to-point to broadcast channel and QMAC. One-shot bound to assess the performance of communication.
%
%\item Strong converse region for quantum broadcast channel. (also quantum MAC).
%
%\item Hypothesis testing converse region.
%
%\item study of some basic channels (to be completed)
%\end{itemize}
%}

%%%%%%%%%%%%%%%%%%%%%%%%%%%%%%%%%%%%%%%%%%%%%%%%%
We explore the classical communication over quantum channels with one sender and two receivers, or with two senders and one receiver, 
%in both the one-shot and asymptotic regime.
%
First, for the quantum broadcast channel (QBC) and the quantum multi-access channel (QMAC), we study the classical communication assisted by non-signalling and positive-partial-transpose-preserving codes, and obtain efficiently computable one-shot bounds to assess the performance of classical communication. 
Second, we consider the asymptotic communication capability of communication over the QBC and QMAC. We derive an efficiently computable strong converse bound for the capacity region, which behaves better than the previous semidefinite programming strong converse bound for point-to-point channels.
Third, we obtain a converse bound on the one-shot capacity region based on the hypothesis testing divergence between the given channel and a certain class of subchannels.
As applications, we analyze the communication performance for some basic network channels, including the classical broadcast channels and a specific class of quantum broadcast channels.
\end{abstract}

% \tableofcontents

%<*body> ********************************************************************************

\section{Introduction}

% The fundamental limit of reliable communication over a noisy quantum channel is characterized by its capacity.

% The quantum communication theory is mainly concerned with the reliable transmission of classical and quantum information over quantum channels. The determination of the fundamental limit of the transmition capacity remains widely unknown, in stark contrast with the study on classical regime, where the capacities can be characterised by simple single-letter or multi-letter formulas.

% classical capacity thm for classical case

% Yard-Hayden Winter works

% One shot case

A fundamental goal of quantum information theory is to find the ultimate limits imposed on information processing and transmission by the laws of quantum mechanics. The classical capacity of a noisy point-to-point quantum channel is the maximal rate at which it can transmit classical information faithfully over asymptotically many uses of the channel. The Holevo-Schumacher-Westmoreland theorem \cite{holevo1973bounds,holevo1998capacity,schumacher1997sending} gives a characterization of the classical capacity of a general quantum channel. 

% broadcast channel
In many situations of communication, there are usually more than one sender and receiver. It is also important to understand how well we could send information via network channels (e.g., broadcast channel, multiple-access channel, interference channel). The capacity region of a classical degraded broadcast channel can be written as a single-letter formula \cite{bergmans1973random,gallager1974capacity}, while no such characterization is known for general classical broadcast channels. Quantum broadcast channel is introduced in \cite{yard2011quantum}, and many information-theoretic results are now known for quantum broadcast channels. Refs. \cite{yard2011quantum,savov2015classical} give a quantum generalization of the superposition coding method for classical communication over quantum broadcast channel. Ref. \cite{radhakrishnan2016one} gives the one-shot Marton inner bound for classical-quantum broadcast channels. A single-letter capacity region of the Hadamard quantum broadcast channel, as a quantum generalization of the degraded broadcast channel, is obtained in \cite{wang2017hadamard}. Refs. \cite{guha2007classical, dupuis2010father, hirche2015improved, seshadreesan2016bounds} offer many other results on communication-related tasks for quantum broadcast channels.

% multi-access channel
The single-letter characterization of the capacity region of classical multi-access channels is given in \cite{ahlswede1973multi,ahlswede1974capacity,liao1972multiple}. Ref. \cite{winter2001capacity} shows the capacity region for classical-quantum multiple access channels admits a single letter characterization. Ref. \cite{hsieh2008entanglement} gives a regularized formula for the entanglement-assisted classical-classical capacity region for quantum multiple access channels, and Ref. \cite{yard2008capacity} studies the classical-quantum and quantum-quantum capacity region.

The capacity of a channel provides a fundamental characterization of the asymptotic information transmission capabilities of the channel, since it is assumed that the senders are allowed to use the channel many times and the channel has no memory after each use. In practice, however, the sender may be forced to use the channel only once and the channel may not be memoryless, and one might be concerned with the tradeoff between the number of channel uses (code blocklength), communication rate and error probability. This is one of the driving forces behind the emerging field of one-shot information theory. In recent years, the quantum broadcast communication protocols were studied in \cite{anshu2017one} using the powerful convex split technique proposed in \cite{anshu2017quantum}, and also studied in \cite{dupuis2010decoupling} using decoupling approach.

% but does not characterize some practically important properties of channels, such as the trade-off 

Another challenge for further study of the quantum channel capacity region is that it is rather difficult to calculate the regularized expression of the capacity region as well as a reasonable estimation, and that little is now known about the strong converse property of quantum network channel. In order to deal with these issues, one can consider the encoding scheme and the decoding scheme as a whole multi-bipartite operation, and impose suitable constraints on this operation, such as non-signalling (NS) operation, positive-partial-transpose-preserving (PPT) operation, and product operation \cite{leung2015power,wang2017semidefinite}.

% study the performance of extra resources in the coding scheme. The extra resource can be PPT operation, non-signalling (NS) operation and shared entanglement. (\XW{To be revised.})

We consider two basic kinds of quantum network channels: the quantum broadcast channel with one sender and two receivers, and the quantum multi-access channel with two senders and one receiver. The motivation for this paper is two-fold. One is to offer an efficiently computable converse bound for classical communication over quantum network channels and provide insights into the study on the strong converse property for general network channels. The other is to study the performance of non-signalling- and PPT- codes in the network communication. In addition, our results on quantum channels may also shed lights on the study of communication over classical network channels.

The paper is structured as follows. In Section \ref{sec:prelim}, we review some relevant preliminaries and set the notations. In Section \ref{sec:broadcast}, after deriving the one-shot communication fidelity of quantum broadcast channel assisted by non-signalling and PPT codes and its classical version, we give an semidefinite-programming strong converse bound on the asymptotic rate region which can be efficiently computed. We also give a hypothesis testing converse bound on the finite-blocklength communication rate. We investigate the non-signalling and PPT-assisted communication fidelity and strong converse bound for the quantum multi-access channel in Section \ref{sec:mac}, and draw a conclusion in Section \ref{sec:last}. 
% \WX{summary of results}

\section{Preliminaries}\label{sec:prelim}

A quantum register $A$ is associated to a Hilbert space $\Hc_A$ equipped with a standard orthonormal basis $\{\ket{j}_A\}_j$. In this work, we only deal with finite-dimensional spaces, and the dimensions of systems $A,B,C$ are denoted by $d_A,d_B,d_C$ respectively. The linear operators from $\Hc_A$ to $\Hc_B$ are always written with subscripts indicating the systems involved, for example, $X_{A\to B}$ and $Y_A$. The subscripts would be omitted when it is clear from context.

%We denote $\Sc (A)$ as the set of density operators \cite{nielsen2002quantum} on system $A$.

A quantum operation (or channel) $\cE_{A\to B}$ with input system $A$ and output system $B$ is a completely positive (CP), trace preserving (TP) linear map from the linear operators on $\Hc_A$ to the linear operators on $B$. Since the subsript of an operator or operation indicates its input and output systems, we can write a product of operators or operations without the tensor symbol, and omit the identity operator $\1$ or identity operation $\id$, which would make no confusion. For example, $X_AY_B\equiv Y_BX_A\equiv X_A\ot Y_B$, $X_{AB}Y_{BC}\equiv (X_{AB}\ot \1_C)(\1_A\ot Y_{BC})$ and $\cE_{B\to C}(X_{AB})\equiv (\id_A\ot\cE_{B\to C})X_{AB}$. We also write the partial trace of a multipartite operator by omitting the subscript the partial trace takes on, for example, $X_B:=\Tr_A(X_{AB})$. The Choi-Jamio{\l}kowski matrix (or Choi matrix for short) \cite{jamiolkowski1972linear,choi1975completely} of a quantum operation $\cE_{A\to B}$ is $J_\cE=\sum_{i,j=1}^{d_A}\ketb{i}{j}\ot\cE(\ketb{i}{j})$ where $\{\ket{i}\}$ is the standard basis of the input space $\cH_A$. The Choi matrix can be equivalently written as $J_\cE=(\id_{\tilde A\to A}\ot\cE_{A\to B})\psi_{\tilde A A}$ where $\psi_{\tilde AA}=\sum_{i,j=1}^{d_A} \ketb{i}{j}_{\tilde A} \ot \ketb{i}{j}_A$ is the unnormalised isotropic maximally entangled state and $\tilde A,A$ are isomorphic systems. The output of the channel $\cE_{A\to B}$ with input $\rho_A$ can be recovered from $J_\cE$ as $\cE_{A\to B}(\rho_A)=\Tr_A(J_\cE\PT{A}\rho_A)$, where $\s{T}_{\!A}$ denotes the partial transpose on $A$. Given an operator $X_{A\to B}=\sum_{ij}x_{ij}\ketb{i_B}{j_A}$, define $\vc (X)=\sum_{ij}x_{ij}\ket{j_A}\ket{i_B}\in\cH_{AB}$. Throughout this paper, log denotes the binary logarithm.

A positive semidefinite operator $P_{AB}$ is said to be a positive partial transpose (PPT) operator if $P\PT{A}\ge 0$. A bipartite operation $\cZ_{AB\to A'B'}$ is PPT-preserving \cite{rains2001semidefinite,rains1999bound} if it takes any PPT density operator to another PPT state. A bipartite operation $\cZ_{AB\to A'B'}$ is non-signalling from $A$ to $B$ if $\Tr_{A'}\cZ_{AB\to A'B'}=\cZ_{B\to B'}\Tr_A$ for some operation $\cZ_{B\to B'}$, and $\cZ_{AB\to A'B'}$ is non-signalling from $B$ to $A$ if $\Tr_{B'}\cZ_{AB\to A'B'}=\cZ_{A\to A'}\Tr_B$ for some operation $\cZ_{A\to A'}$. One-way non-signalling operations are also referred to
as `semi-causal' \cite{beckman2001causal,eggeling2002semicausal}.

%See \cite{nielsen2011quantum} for more introduction on quantum information theory.

% Throughout this paper we use $\phi$ to denote normalized maximally entangled state.

% \section{Quantum multipartite no-signalling operation}
% In \cite{abbott2016multipartite}, the classical multipartite causal correlations are defined as follows. For $n=1$, any valid probability distribution $p(a_1,x_1)$ is causal. For $n\ge 2$, an $n$-partite correlation is causal iff it can be written in the form $p(\vec a|\vec x)=\sum_{k\in[n]} q_k p_k(a_k|x_k) p_{k,x_k,a_k}(\vec a_{\backslash k} | \vec x_{\backslash k})$, with $q_k\ge 0,\sum_k q_k=1$, where $p_k(a_k|x_k)$ and $p_{k,x_k,a_k}(\vec a_{\backslash k} | \vec x_{\backslash k})$ are single-party and $(n-1)$-partite correlation.

% \section{Classical capacity of 1-to-1 channel}
% Let $\cN_{A'\to B}$ be a channel and $\Om$ the set of NSPPT codes $\cX_{AB\to A'B'}$. The overall channel is $\cM_{A\to B'}=\cN_{A'\to B}\circ\cX_{AB\to A'B'}$.

% the optimal success probability is $f_\Om(\cN,m):=\max_{\cX\in\Om} \f{1}{m}\sum_{k=1}^m \brak{ \cM(\ketb{k}{k}), \ketb{k}{k} } = \f{1}{m}\max_\cX \Tr( J_{\cM,AB'} D_{AB'} )$ for $D=\sum_{k=1}^m \ketb{k}{k}\ot \ketb{k}{k}$.

A \emph{code} in our network communcation protocol is defined as some tripartite operation $\cX$. We say a tripartite operation $\cX_{ABC\to A'B'C'}$ is non-signalling (NS) and positive-partial-transpose-preserving (PPT) if and only if it is NS and PPT with respect to any bipartite cut. The condition is that its Choi matrix $X_{ABCA'B'C'}$ satisfies \cite{leung2015power,duan2016no}
\Als{\label{NSPPT-condition}
\x{CP}\quad & X_{ABCA'B'C'}\ge 0, \\
\x{TP}\quad & X_{ABC}=\1_{ABC}, \\
\x{PPT}\quad & X\PT{AA'}\ge0, X\PT{BB'}\ge0, X\PT{CC'}\ge0, \\
A\not\leftrightarrow BC \quad & X_{ABCB'C'}=\f{\1_A}{d_A}\ot X_{BCB'C'}, X_{ABCA'}=\f{\1_{BC}}{d_{BC}}\ot X_{AA'}, \\
B\not\leftrightarrow AC \quad & X_{ABCA'C'}=\f{\1_B}{d_B}\ot X_{ACA'C'}, X_{ABCB'}=\f{\1_{AC}}{d_{AC}}\ot X_{BB'}, \\
C\not\leftrightarrow AB \quad & X_{ABCA'B'}=\f{\1_C}{d_C}\ot X_{ABA'B'}, X_{ABCC'}=\f{\1_{AB}}{d_{AB}}\ot X_{CC'}.
}

The unassisted code corresponds to some product tripartite operation $\cX = \cE_{A\to A'}\cD_{1,B\to B'} \cD_{2,C\to C'}$. A \emph{code class} $\Om$ is a set of codes satisfying certain properties. The set of NS and PPT tripartite operations, the set of NS tripartite operations, and the set of product operations are written as $\Om=\x{NSPPT}$, $\Om=\x{NS}$ and $\Om=\x{ua}$, respectively.

Semidefinite programming (SDP), as a generalization of linear programming, has been proven to be a 
a very useful tool in the theory of quantum information and computation (see, e.g., \cite{leung2015power,wang2017semidefinite,matthews2014finite,jain2010qip,wang2017irreversibility,xie2017approximate} for a partial list.)
SDP can be solved via interior point methods efficiently in theory as well as in practice.
% SDP is solvable via interior point methods and it can be solved very efficiently in practice as well as in theory.
In this paper, we use the CVX software \cite{Grant2008a} and QETLAB \cite{NathanielJohnston2016} to solve SDPs.

\section{Classical communication over quantum broadcast channel}\label{sec:broadcast}

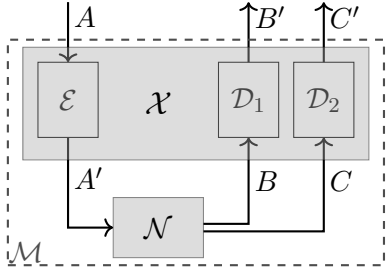
\begin{figure}[htbp]
\centering
\begin{tikzpicture}
    \def\step{0.2cm}  \def\xo{0.06cm}
    % \draw[help lines,dashed,step=\step,xshift=0cm, yshift=0cm] (0,0) grid (4.6,2.8);
    \draw[black,fill=black!33,opacity=0.4] (1.2,0) rectangle (2.4,0.8) node[midway,black,opacity=1] {$\cN$};
    \draw (0.2,1.6) rectangle (1,2.6) node[midway] {$\cE$};
    \draw (2.6,1.6) rectangle (3.4,2.6) node[midway] {$\cD_1$};
    \draw (3.6,1.6) rectangle (4.4,2.6) node[midway] {$\cD_2$};
    \draw[thick,->] (0.6,3.4) -- node[right,xshift=-\xo,yshift=\step] {$A $} (0.6,2.6);
    \draw[thick,<-] (3  ,3.4) -- node[right,xshift=-\xo,yshift=\step] {$B'$} (3,2.6);
    \draw[thick,<-] (4  ,3.4) -- node[right,xshift=-\xo,yshift=\step] {$C'$} (4,2.6);
    \draw[thick,->] (0.6,1.6) -- node[right,xshift=-\xo,yshift=\xo] {$A'$} (0.6,0.4) -- (1.2,0.4);
    \draw[thick,->] (2.4,0.45) -- (3,0.45) -- node[right,xshift=-\xo] {$B$} (3,1.6);
    \draw[thick,->] (2.4,0.35) -- (4,0.35) -- node[right,xshift=-\xo,yshift=0.05cm] {$C$} (4,1.6);
    \draw[black,fill=black!33,opacity=0.4] (0,1.3) rectangle (4.6,2.8) node[midway,black,opacity=1,xshift=-0.5cm] {\large $\cX$};

    % \node [above] at (3,3.4) {\scriptsize $\widehat x_1$};
    % \node [above,xshift=\step,yshift=0.45cm,rotate=90] at (3,3.4) {\tiny $\in$};
    % \node [above,xshift=0cm,yshift=0.45cm] at (3,3.4) {\tiny $[m_1]$};
    % \node [above] at (4,3.4) {\scriptsize $\widehat x_2$};
    % \node [above,xshift=\step,yshift=0.45cm,rotate=90] at (4,3.4) {\tiny $\in$};
    % \node [above,xshift=0cm,yshift=0.45cm] at (4,3.4) {\tiny $[m_2]$};
    % \node [above] at (0.6,3.4) {\scriptsize $(x_1,x_2)$};
    % \node [above,xshift=\step,yshift=0.45cm,rotate=90] at (0.6,3.4) {\tiny $\in$};
    % \node [above,xshift=0cm,yshift=0.45cm] at (0.6,3.4) {\tiny $[m_1]\times [m_2]$};
    \draw[thick,dashed,black!70] (-0.2,-0.1) node[xshift=0.24cm,yshift=0.17cm] {$\cM$} rectangle (4.8,2.9);
\end{tikzpicture}
\caption{Classical communication over quantum broadcast channel $\cN$ assisted by a code $\cX$. The tripartite code $\cX$ is designed in order for the whole operation $\cM$ to emulate a noiseless classical channel.}
\label{fig:1-to-2}
\end{figure}

Suppose Alice wants to send classical message labeled by $\{1,\dots,m_1\}$ to Bob, and simultaneously send message labeled by $\{1,\dots,m_2\}$ to Charlie, using the composite channel $\cM=\cN\circ\cX$, where $\cX$ is a tripartite operation as a coding scheme; see Fig. \ref{fig:1-to-2}. The code $\cX$ is chosen within some coding class $\Om$ in order to make the overall channel $\cM$ as close to a classical noiseless channel as possible. Thus the registers $A,B',C'$ can be assumed to be classical \cite{wang2017semidefinite}. The classical register $A$ indeed consists of two subregisters $A_1$ and $A_2$, storing messages to be sent to Bob and Charlie respectively.

\subsection{One-shot $\epsl$-error capacity}

We first define several quantities to characterize the capability for communication over quantum broadcast channels.

\Def{\label{def1}
    The success probability of $\cN_{A'\to BC}$ to transmit messages of size $(m_1,m_2)$ assisted by code $\cX_{ABC\to A'B'C'}$ is defined as
    \Eq{
        p_s(\cN,\cX,m_1,m_2) = \f{1}{m_1m_2}\sum_{i,j=1}^{m_1,m_2} \Tr(\cM(\ketb{ij}{ij}_A)\ketb{ij}{ij}_{B'C'}),
    }
    where $\cM_{A\to B'C'}=\cN\circ\cX$.

    Moreover, the $\Om$-assisted optimal success probability of $\cN$ to transmit messages of size $(m_1,m_2)$ is defined as
    \Eq{
        f_{\Om}(\cN,m_1,m_2)=\max_{\cX\in\Om} p_s(\cN,\cX,m_1,m_2).
    }
}

\Def{\label{def2}
    The $\Om$-assisted one-shot $\epsl$-error classical capacity region of $\cN$ is defined as 
    \begin{equation}
    \{(R_1,R_2):f_\Om(\cN,2^{R_1},2^{R_2})\ge 1-\epsl\}.
    \end{equation}

    The $\Om$-assisted classical capacity region of $\cN$ is defined as 
    \begin{equation}
    \{(R_1,R_2):\lim_{n\to\infty} f_\Om (\cN^{\ot n},2^{nR_1},2^{nR_2})=1\}.
    \end{equation}

    We call $(R_1,R_2)$ a strong converse rate pair for $\cN$, if
    \Eq{
    \lim_{n\to\infty} f_\x{ua} (\cN^{\ot n},2^{nR_1},2^{nR_2})=0.
    }
    % The $\Om$-assisted strong converse rate region of $\cN$ is defined as
    % \Eq{
    %     \{(R_1,R_2):\}. }
        % \Rc_{\x{sc},\Om}(\cN):=\{(R_1,R_2):\lim_{n\to\infty} f_\Om (\cN^{\ot n},2^{nR_1},2^{nR_2})=0\}. }
    % Denote by $\Rc_\x{c}^{1,\epsl}(\cN)$ and $\Rc_\x{sc}(\cN)$ the unassisted one-shot $\epsl$-error converse rate region of $\cN$ and the unassisted strong converse rate region of $\cN$, respectively.
}

% \Def{\label{def3}
%     The $\Om$-assisted one-shot $\epsl$-error converse rate region of $\cN$ is defined as
%     \Eq{
%         \{(R_1,R_2):f_\Om(\cN,2^{R_1},2^{R_2})\le 1-\epsl\}. }
%         % \Rc_{\x{c},\Om}^{(1),\epsl}(\cN):=\{(R_1,R_2):f_\Om(\cN,2^{R_1},2^{R_2})\le 1-\epsl\}. }
% }

We show that the NSPPT-assisted optimal success probability of a broadcast channel can be formulated as the following SDP.

\Thm{
The optimal success probability of a channel $\cN_{A'\to BC}$ to transmit messages of size $(m_1,m_2)$ assisted by NSPPT codes is given by
\Als{\label{original-sdp-fidelity}
    f_\x{NSPPT} (\cN,m_1,m_2) = \x{max } & \Tr\ps{ J_\cN\T E_{1,A'BC} } \\
    \x{s.t. } & E_i,E_i\PT{B},E_i\PT{C},E_i\PT{A'}\ge 0 \x{ for } i=1,2,3,4, \ (\x{CP, PPT}) \\
    & K_{A'BC} = K_{A'} \f{\1_{BC}}{d_{BC}}, \ (BC\not\to A) \\
    & \Tr K_{A'} =d_{BC}, \ (\x{TP}) \\
    & E_{i,BC}=\f{\1_{BC}}{m_1m_2} \ \x{ for } i=1,2,3, \ (A\not\to BC) \\
    & E_{1,A'BC}+(m_2-1)E_{2,A'BC} = (E_{1,A'B}+(m_2-1)E_{2,A'B}) \ot \1_C/d_C, \\ %\ (C\not\to AB) \\
    % &  E_{3,A'BC}+(m_2-1)E_{4,A'BC} = (E_{3,BA'}+(m_2-1)E_{4,BA'}) \ot\f{\1_C}{d_C}, \ (C\not\to AB) \\
    & E_{1,A'BC}+(m_1-1)E_{3,A'BC} = (E_{1,A'C}+(m_1-1)E_{3,A'C}) \ot \1_B/d_B, %\ (B\not\to AC)
    % &  E_{2,A'BC}+(m_1-1)E_{4,A'BC} = (E_{2,CA'}+(m_1-1)E_{4,CA'}) \ot\f{\1_B}{d_B}. \ (B\not\to AC) 
}
    where $K_{A'BC} := E_{1,A'BC}+(m_2-1)E_{2,A'BC} +(m_1-1)E_{3,A'BC}+(m_1-1)(m_2-1)E_{4,A'BC}$.
% where the last two lines can be equivalently replaced by 
% \Als{
%     & E_{3,A'BC}+(m_2-1)E_{4,A'BC} \\
%     & \phantom{www} = (E_{3,BA'}+(m_2-1)E_{4,BA'}) \ot\f{\1_C}{d_C}, \\
%     & E_{2,A'BC}+(m_1-1)E_{4,A'BC} \\
%     & \phantom{www} = (E_{2,CA'}+(m_1-1)E_{4,CA'}) \ot\f{\1_B}{d_B}.
%     }
}

\begin{proof}

Denoting $\cM_{A\to B'C'}=\cN_{A'\to BC}\circ\cX_{ABC\to A'B'C'}$, $D_{1,A_1B'}=\sum_{i=1}^{m_1}\ketb{ii}{ii}$ and $D_{2,A_2C'}=\sum_{j=1}^{m_2}\ketb{jj}{jj}$, the success probability of $\cN$ to transmit messages of size $(m_1,m_2)$ assisted by code $\cX$ is written as
\Als{
    p_s(\cN,\cX,m_1,m_2) :=&\ \f{1}{m_1m_2}\sum_{i,j=1}^{m_1,m_2} \Tr(\cM(\ketb{ij}{ij}_A)\ketb{ij}{ij}_{B'C'}) \\
    % =&\ \f{1}{m_1m_2} \Tr\ps{ \sum_{i,j,k,l=1}^{m_1,m_2,m_1,m_2} (\ketb{ij}{kl}_{A_1A_2} \ot \cM(\ketb{ij}{kl}_{A_1A_2})) (D_1\ot D_2)} \\
    =&\ \f{1}{m_1m_2} \Tr\ps{ J_{\cM,AB'C'}(D_{1,A_1B'}\ot D_{2,A_2C'}) }
}

Since $J_{\cM}=\Tr_{A'BC}(J_{\cN}\T X_{ABCA'B'C'})$, we have
\Al{
    f_\x{NSPPT} (\cN,m_1,m_2)  &= \f{1}{m_1m_2}\max_X \Tr\ps{ J\T_{\cN,A'BC} X (D_{1,A_1B'} \ot D_{2,A_2C'}) } \\
    &= \f{1}{m_1m_2}\max_{\tilde X} \Tr\ps{ J\T_{\cN,A'BC} \tilde X (D_{1,A_1B'} \ot D_{2,A_2C'}) },
}
where $\tilde X :=\f{1}{m_1!m_2!}\sum_{\pi_1\in S_{m_1},\pi_2\in S_{m_2}} \ps{W_{\pi_1,A_1B'}\ot W_{\pi_2,A_2C'}} X \ps{W_{\pi_1,A_1B'}\ot W_{\pi_2,A_2C'}}\dg $. Here $S_{m_1},S_{m_2}$ are symmetric groups, and $W_{\pi_1}$ and $W_{\pi_2}$ are the operators permuting the basis with respect to $\pi_1$ and $\pi_2$ respectively. If $X$ is a feasible solution to the optimization problem, so is $\tilde X$. Since $X\ge 0$ and the registers $A,B',C'$ are classical, $\tilde X$ can be written as
\Als{\label{eqn5}
    \tilde X =&\ D_{1,A_1B'}\ot D_{2,A_2C'} \ot E_{1,A'BC} + D_{1,A_1B'}\ot (\1-D_{2,A_2C'}) \ot E_{2,A'BC} \\
    &\ +(\1-D_{1,A_1B'})\ot D_{2,A_2C'} \ot E_{3,A'BC} + (\1-D_{1,A_1B'})\ot (\1-D_{2,A_2C'}) \ot E_{4,A'BC}
}
for some positive semidefinite operators $E_i$. The CP and PPT constraints are equivalent to
\Eq{\label{constraint-cp-ppt}
    E_i,E_i\PT{B},E_i\PT{C},E_i\PT{A'}\ge 0,
}
for each $i$. Using Eq. (\ref{eqn5}) leads to a simplification
\Eq{
    f_\x{NSPPT} (\cN,m_1,m_2) = \max_{E_1} \Tr\ps{ J_\cN\T E_{1,A'BC} }.
}

Denoting
\Eq{\label{def-LBCA}
    K_{A'BC}:=E_{1,A'BC}+(m_2-1)E_{2,A'BC}+(m_1-1)E_{3,A'BC}+(m_1-1)(m_2-1)E_{4,A'BC},
}
the TP constraint $\tilde X_{ABC}=\1_{ABC}$ is equivalent to
\Eq{\label{constraint-tp1}
    K_{BC}=\1_{BC}.
}

The NS constraint $BC\not\to A$ is equivalent to
\Eq{\label{NS-BC-A}
K_{A'BC}=K_{A'} \f{\1_{BC}}{d_{BC}}.
}
Using this constraint, Eq. (\ref{constraint-tp1}) becomes
\Eq{\label{constraint-tp}
    \Tr K_{A'}=d_{BC}.
}
% If we denote $\rho_{A'}:=\f{L_{A'}}{d_{BC}}$, then the constraints (\ref{NS-BC-A}) and (\ref{constraint-tp}) are equivalent to $L_{A'BC}=\1_{BC}\rho_{A'}$ and $\Tr\rho_{A'}=1$ respectively.

The NS constraint $A\not\to BC$ is equivalent to $D_1D_2 E_{1,BC}+D_1(\1-D_2) E_{2,BC}+(\1-D_1)D_2 E_{3,BC}+(\1-D_1)(\1-D_2) E_{4,BC}=\f{1}{d_A}\1_{ABCB'C'}=\f{1}{m_1m_2}\1_{ABCB'C'}$. It follows that
\Eq{\label{NS-A-BC}
    E_{i,BC}=\f{\1_{BC}}{m_1m_2} \x{ for } i=1,2,3,4.
}

Since $K_{BC}=\1_{BC}$, that $E_{4,BC}=\f{\1_{BC}}{m_1m_2}$ follows from that $E_{i,BC}=\f{\1_{BC}}{m_1m_2}$, $i=1,2,3$. Since $\tilde X_{ABCC'}=\f{1}{m_2}\1_{ABCC'}$ and $\tilde X_{ABCB'}=\f{1}{m_1}\1_{ABCB'}$, the NS constraint $AB\not\to C$ and $AC\not\to B$ are satisfied.

The constraint $C\not\to AB$ is
\Als{
    & D_1E_{1,A'BC}+(m_2-1)D_1E_{2,A'BC}+(\1-D_1)E_{3,A'BC}+(m_2-1)(\1-D_1)E_{4,A'BC} \\
    & = \ps{ D_1E_{1,A'B}+(m_2-1)D_1E_{2,A'B}+(\1-D_1)E_{3,A'B}+(m_2-1)(\1-D_1)E_{4,A'B} } \ot\f{\1_C}{d_C},
}
which is equivalent to
\Al{
    E_{1,A'BC}+(m_2-1)E_{2,A'BC} &= (E_{1,A'B}+(m_2-1)E_{2,A'B}) \ot\f{\1_C}{d_C}, \label{NS-C-AB-1}\\
    E_{3,A'BC}+(m_2-1)E_{4,A'BC} &= (E_{3,A'B}+(m_2-1)E_{4,A'B}) \ot\f{\1_C}{d_C} \label{NS-C-AB-2}.
}

Similarly, the constraint $B\not\to AC$ is equivalent to
\Al{
    E_{1,A'BC}+(m_1-1)E_{3,A'BC} &= (E_{1,A'C}+(m_1-1)E_{3,A'C}) \ot\f{\1_B}{d_B}, \label{NS-B-AC-1}\\
    E_{2,A'BC}+(m_1-1)E_{4,A'BC} &= (E_{2,A'C}+(m_1-1)E_{4,A'C}) \ot\f{\1_B}{d_B} \label{NS-B-AC-2}.
}

Notice that Eq. (\ref{NS-C-AB-2}) is also implied by Eqs. (\ref{NS-BC-A}) and (\ref{NS-C-AB-1}), and that Eq. (\ref{NS-B-AC-2}) is also implied by Eqs. (\ref{NS-BC-A}) and (\ref{NS-B-AC-1}).

Putting together the above constraints, we obtain the desired SDP characterization.
 % (\ref{constraint-cp-ppt}), (\ref{NS-BC-A}), (\ref{constraint-tp}), (\ref{NS-A-BC}), (\ref{NS-C-AB-1}), (\ref{NS-C-AB-2}), (\ref{NS-B-AC-1}) and (\ref{NS-B-AC-2}), with notation (\ref{def-LBCA})
\end{proof}

\subsection{Reduction to classical case}
For a c-qq channel $\cN_{A'\to BC}$ with Choi matrix $J_\cN=\sum_{ij}\ketb{ij}{ij}_{A'}\ot\rho_{i,B}\ot\sg_{j,C}$, its optimal success probability can be simplified as follows.

Let $E_k=\sum_{ij}\ketb{ij}{ij}_{A'}\ot F_{k,ij,BC}$, then $K_{A'BC}=\sum_{ij}\ketb{ij}{ij}\ot( F_{1,ij}+(m_2-1)F_{2,ij}+(m_1-1)F_{3,ij}+(m_1-1)(m_2-1)F_{4,ij} ) =:\sum_{ij}\ketb{ij}{ij}\ot T_{ij}$. Thus $K_{A'}=\sum_{ij}\ketb{ij}{ij}\Tr(T_{ij})$. The constraint $BC\not\to A$ becomes $T_{ij}=\Tr(T_{ij})\f{\1_{BC}}{d_{BC}}$ for each $i,j$. The TP constraint becomes $\sum_{ij}\Tr(T_{ij})=d_{BC}$. The constraint $A\not\to BC$ becomes $\sum_{ij}F_{k,ij}=\f{\1_{BC}}{m_1m_2}$ for each $k$. The constraint $C\not\to AB$ becomes $F_{1,ij,BC}+(m_2-1)F_{2,ij,BC} = (F_{1,ij,B}+(m_2-1)F_{2,ij,B})\ot\f{\1_C}{d_C}$ for each $i,j$. The constraint $B\not\to AC$ becomes $F_{1,ij,BC}+(m_1-1)F_{3,ij,BC} = (F_{1,ij,C}+(m_1-1)F_{3,ij,C})\ot\f{\1_B}{d_B}$ for each $i,j$.
\
% \WX{Not sure whether the collective povm on BC performs better than local povm}

\Prop{
Given a c-qq channel $\cN_{A'\to BC}$ which outputs $\rho_{i,B}$ and $\sg_{j,C}$ upon input $i,j$. The optimal success probability of the channel $\cN_{A'\to BC}$ to transmit messages of size $(m_1,m_2)$ assisted by NSPPT codes is given by
\Als{\label{original-sdp-fidelity-cqq}
    f_\x{NSPPT} & (\cN,m_1,m_2) \\
    = \x{max } & \sum_{ij}\Tr(F_{1,ij}(\rho_i\T\ot\sg_j\T)) \\
    \x{s.t. } & F_{k,ij},F_{k,ij}\PT{B} \ge 0 \ \forall k,i,j, \ (\x{CP, PPT}) \\
    & T_{ij}=\Tr(T_{ij})\f{\1_{BC}}{d_{BC}}, \ (BC\not\to A) \\
    & \sum_{ij}\Tr(T_{ij})=d_{BC}, \ (\x{TP}) \\
    & \sum_{ij}F_{k,ij}=\f{\1_{BC}}{m_1m_2} \ \forall k, \ (A\not\to BC) \\
    & F_{1,ij,BC}+(m_2-1)F_{2,ij,BC} = (F_{1,ij,B}+(m_2-1)F_{2,ij,B})\ot \1_C/d_C \ \forall i,j, \ (C\not\to AB) \\
    & F_{1,ij,BC}+(m_1-1)F_{3,ij,BC} = (F_{1,ij,C}+(m_1-1)F_{3,ij,C})\ot \1_B/d_B, \ \forall i,j, \ (B\not\to AC)
}
where $T_{ij}=F_{1,ij}+(m_2-1)F_{2,ij}+(m_1-1)F_{3,ij}+(m_1-1)(m_2-1)F_{4,ij}$.
}

Furthermore, we can reduce to the case of classical channel. When $\rho_i,\sg_j$ are all classical states, if $\{F_{k,ij}\}$ is feasible solution to the above SDP, so is $\{\Dt(F_{k,ij})\}$, where $\Dt$ is completely dephasing channel. Thus it suffices to consider the case that $F_{k,ij}$ are classical for a classical channel $\cN$. Suppose a classical channel output $y=i',z=j'$ upon input $x=(i,j)$ with probability $p(i'j'|ij)$, and its Choi matrix is $\sum_{iji'j'} \ketb{ij}{ij}\ot p(i'j'|ij)\ketb{i'j'}{i'j'}$. Let $E_k=\sum_{iji'j'}\ketb{ij}{ij}\ot \eta_k(i'j'ij)\ketb{i'j'}{i'j'}$. In this case, the SDP reduces to a linear programming and the PPT constraints are implicitly satisfied.

\Thm{\label{classical-case}
The optimal success probability of a classical channel $p(i'j'|ij)$ to transmit messages of size $(m_1,m_2)$ assisted by NS codes is given by
\Als{\label{original-sdp-fidelity-ccc}
    f_\x{NS} & (\cN,m_1,m_2) \\
    = \x{max } & \sum_{iji'j'} p(i'j'|ij) \eta_1(i'j'ij) \\
    \x{s.t. }  & \eta_k(i'j'ij)\ge 0,\ \forall k,i',j',i,j, \\
    & \mu(i'j'ij)=\eta_1(i'j'ij)+(m_2-1) \eta_2(i'j'ij)  \\
    & \phantom{\mu(i'j'ij)=}      +(m_1-1) \eta_3(i'j'ij)+(m_1-1)(m_2-1) \eta_4(i'j'ij),  \\
    & \mu(ij)=d_{BC} \mu(i'j'ij) \ \forall i',j', \ (BC\not\to A) \\
    & \sum_{ij} \mu(ij)=d_{BC},  \\
    & m_1m_2 \eta_k(i'j')=1 \ \forall k, \ (A\not\to BC) \\
    & d_C (\eta_1(i'j'ij)+(m_2-1)\eta_2(i'j'ij))= \eta_1(i'ij)+(m_2-1) \eta_2(i'ij),\ \forall i',j',i,j, \ (C\not\to AB) \\
    & d_B (\eta_1(i'j'ij)+(m_1-1)\eta_3(i'j'ij))= \eta_1(j'ij)+(m_1-1) \eta_3(j'ij),\ \forall i',j',i,j, \ (B\not\to AC)
}
where $\mu(ij):=\sum_{i'j'} \mu(i'j'ij)$, $\eta_k(i'j'):=\sum_{ij}\eta_k(i'j'ij)$.
}

Theorem \ref{classical-case} gives a linear-programming converse bound on the optimal success probability of unassisted communication over classical channel, and can be viewed as an extension of the works of \cite{polyanskiy2010channel,matthews2012linear}. For the case of point-to-point classical communication, Polyanskiy, Poor and Verd{\'u} prove a general converse bound on the one-shot $\epsl$-error capacity \cite{polyanskiy2010channel}, and then Matthews gives a linear programming characterization for this converse bound via non-signalling code \cite{matthews2012linear}. Here we present a linear programming converse bound for classical broadcast communication. It remains for future work to compare this bound with some existing results, e.g., \cite{nair2007outer}.

% It remains to give the capacity region of a classical channel using this formula of optimal success probability.

% \WX{ Yes, the constraints on real-value variables $Q_{k,i'j'ij}$ are linear}

% \WX{no PPV-like linear-programming converse bound for classical broadcast communication found}

\subsection{Converse bound on classical capacity region based on NSPPT codes}
% Due to \ref{original-sdp-fidelity}, we have
% \Als{\label{sdp-fidelity-conv}
%     f_\x{NSPPT} & (\cN,m_1,m_2) \\
%     \le \x{max } & \Tr\ps{ J_\cN\T E_{1,A'BC} } \\
%     \x{s.t. } & E_1,E_1\PT{B},E_1\PT{C} \ge 0, \\
%     & E_{1,BC}=\f{\1_{BC}}{m_1m_2}, \\
%     & \Tr E_1+(m_2-1)\Tr E_2+(m_1-1)\Tr E_3\led_{BC}, \\
%     & E_{1,A'BC}+(m_2-1)E_{2,A'BC} = (E_{1,A'B}+(m_2-1)E_{2,A'B}) \ot\f{\1_C}{d_C},\\
%     & E_{1,A'BC}+(m_1-1)E_{3,A'BC} = (E_{1,A'C}+(m_1-1)E_{3,A'C}) \ot\f{\1_B}{d_B}.
% }

% The Lagrange of this SDP is 
% \Aln{
%     \x{Lag} = & \Tr\ps{ J_\cN\T E_{1,A'BC} } + \Tr(E_1\PT{B} P_1) + \Tr(E_1\PT{C} P_2) \\
%     & + \Tr((\1_{BC}-m_1m_2E_{1,BC})Q_{BC}) \\
%     & + (d_{BC} -\Tr E_1-(m_2-1)\Tr E_2-(m_1-1)\Tr E_3 ) x \\
%     & + \Tr \bigg( \bigg( E_{1,A'BC}+(m_2-1)E_{2,A'BC} - (E_{1,A'B}+(m_2-1)E_{2,A'B}) \ot\f{\1_C}{d_C} \bigg)R_1 \bigg) \\
%     & + \Tr \bigg( \bigg( E_{1,A'BC}+(m_1-1)E_{3,A'BC} - (E_{1,A'C}+(m_1-1)E_{3,A'C}) \ot\f{\1_B}{d_B} \bigg)R_2 \bigg) \\
%     =& \Tr\bigg(E_1\big( J_\cN\T+P_1\PT{B}+P_2\PT{C}-m_1m_2\1_A Q_{BC}+R_{1,A'BC}-R_{1,A'B}\f{\1_C}{d_C}+R_{2,A'BC}-R_{2,A'C}\f{\1_B}{d_B}-x \big)\bigg) \\
%     & + \Tr\bigg(  E_2\big( (m_2-1)(R_{1,A'BC}-R_{1,A'B} \1_C/d_C)-x \big)  \bigg) \\
%     & + \Tr\bigg(  E_3\big( (m_1-1)(R_{2,A'BC}-R_{2,A'C} \1_B/d_B)-x \big)  \bigg) \\
%     & + \Tr Q_{BC} +xd_{BC}
% }
% where $Q,R_i\in\x{Herm}$ and $x,P_i\ge 0$.

Before introducing the strong converse for the broadcast capacity, we first present an SDP upper bound on the single-shot optimal success probability of NSPPT codes.

\Prop{\label{prop:g(N)}
    For any quantum broadcast channel $\cN_{A' \to BC}$ and given $m_1,m_2$,
    \Eq{
    f_\x{NSPPT}(\cN,m_1,m_2) \le g(\cN,m_1,m_2),
    }
    where
    \Als{
        \label{sdp-fidelity-conv-dual2}
        g(\cN,m_1,m_2) := \x{min } & \Tr Q_{BC} \\
        \x{s.t. } & m_1m_2\1_{A'} Q_{BC} \ge V_{A'BC} \ge -m_1m_2\1_{A'}Q_{BC}, \\
        & V_{A'BC}\PT{C} \ge Y_{A'BC}\PT{C} \ge -V_{A'BC}\PT{C}, \\
        & Y_{A'BC}\PT{B} \ge Z_{A'BC}\PT{B} \ge -Y_{A'BC}\PT{B}, \\
        & Z_{A'BC}\PT{A'} \ge J_{\cN}\PT{BC} \ge -Z_{A'BC}\PT{A'}.
    }
    
    Furthermore, $g$ is submultiplicative in the sense that 
    \begin{equation}
g(\cN\ot\cN',m_1m_1',m_2m_2') \le g(\cN,m_1,m_2)g(\cN',m_1',m_2')
\end{equation}
 Consequently, $f_\x{NSPPT}(\cN^{\ot n},m_1^n,m_2^n)\le g(\cN,m_1,m_2)^n$.
}

\begin{proof}
It follows from Eq. (\ref{original-sdp-fidelity}) that
\Als{\label{sdp-fidelity-conv}
    f_\x{NSPPT}  (\cN,m_1,m_2) \le \x{max } & \Tr\ps{ J_\cN\T E_{1,A'BC} } \\
    \x{s.t. } & E_1,E_1\PT{A'},E_1\PT{B},E_1\PT{C} \ge 0, \\
    & E_{1,BC}=\f{\1_{BC}}{m_1m_2}.
    }

% The Lagrange of this SDP is 
% \Aln{
%     & \Tr\ps{ J_\cN\T E_{1,A'BC} } + \Tr(E_1\PT{A'} P_1) +\Tr(E_1\PT{B} P_2) + \Tr(E_1\PT{C} P_3) 
%       + \Tr((\1_{BC}-m_1m_2E_{1,BC})Q_{BC})\\
%     & = \Tr(E_1(J_\cN\T +P_1\PT{A'}+P_2\PT{B}+P_3\PT{C}-m_1m_2\1_AQ_{BC})) +\Tr(Q_{BC}),
% }
% where $P_i\ge 0,Q=Q\dg$.

The dual of SDP (\ref{sdp-fidelity-conv}) is
\Als{\label{sdp-fidelity-conv-dual1}
    \x{min } & \Tr Q_{BC} \\
    \x{s.t. } & J_\cN\T +P_1\PT{A'}+P_2\PT{B}+P_3\PT{C}-m_1m_2\1_{A'}Q_{BC} \le 0 \\
    & P_i\ge 0.
    }

Here the Hermicity of $Q_{BC}$ is implicitly implied. Noticing that $\1_{A'BC}/d_{A'BC}>0$ is a feasible solution to SDP (\ref{sdp-fidelity-conv}), the optimal solutions to SDPs (\ref{sdp-fidelity-conv}) and (\ref{sdp-fidelity-conv-dual1}) coincide due to the Slater's theorem.

% \XW{Strong duality?
% The dual of this SDP is
% \Als{\label{sdp-fidelity-conv-dual1}
%     \x{min } & \Tr Q_{BC} \\
%     \x{s.t. } & J_\cN +P_1\PT{A'}+P_2\PT{B}+P_3\PT{C}-m_1m_2\1_AQ_{BC} \le 0 \\
%     & P_i\ge 0, Q=Q\dg.
%     }
% }

% \XW{More details about ``rewritten''}

Denoting $V:=J_\cN\T +P_1\PT{A'}+P_2\PT{B}+P_3\PT{C}$, $Y:=J_\cN\T+P_1\PT{A'}+P_2\PT{B}$ and $Z:=J_\cN\T+P_1\PT{A'}$, it follows that the first constraint in (\ref{sdp-fidelity-conv-dual1}) is equivalent to $m_1m_2\1_{A'}Q_{BC}\ge V$, that $P_1\ge 0$ iff $V\PT{C}\ge Y\PT{C}$, that $P_2\ge 0$ iff $Y\PT{B}\ge Z\PT{B}$, and that $P_1\ge 0$ iff $Z\PT{A'}\ge J_\cN\PT{BC}$. SDP (\ref{sdp-fidelity-conv-dual1}) can be rewritten as
\Als{
    \x{min } & \Tr Q_{BC} \\
    \x{s.t. } & m_1m_2\1_{A'}Q_{BC} \ge V_{A'BC}, \\
    & V_{A'BC}\PT{C} \ge Y_{A'BC}\PT{C}, \\
    & Y_{A'BC}\PT{B} \ge Z_{A'BC}\PT{B}, \\
    & Z_{A'BC}\PT{A'} \ge J_{\cN}\PT{BC}.
    }

By adding new constraints, the above SDP is no larger than $g(\cN,m_1,m_2)$.

It can be verified that $g$ is sub-multiplicative, since if $(Q,V,Y,Z)$ and $(Q',V',Y',Z')$ are feasible solutions to $g(\cN,m_1,m_2)$ and $g(\cN',m_1',m_2')$ respectively, then $(Q\ot Q',V\ot V',Y\ot Y',Z\ot Z')$ is feasible to $g(\cN\ot\cN',m_1 m_1',m_2 m_2')$.
\end{proof}

\Thm{\label{thm:converse}
    For a quantum broadcast channel $\cN_{A'\to BC}$, if $R_1+R_2 > C_g(\cN)$ then $(R_1,R_2)$ is a strong converse rate pair. Here
    \Als{\label{sdp-fidelity-conv-dual-thm}
    C_g(\cN) := \log \x{min } & \Tr Q_{BC} \\
    \x{s.t. } & \1_{A'}Q_{BC} \ge V_{A'BC} \ge -\1_{A'}Q_{BC} \\
    & V_{A'BC}\PT{C} \ge Y_{A'BC}\PT{C} \ge -V_{A'BC}\PT{C} \\
    & Y_{A'BC}\PT{B} \ge Z_{A'BC}\PT{B} \ge -Y_{A'BC}\PT{B} \\
    & Z_{A'BC}\PT{A'} \ge J_{\cN}\PT{BC} \ge -Z_{A'BC}\PT{A'}.
    }
    % Then $\Rc(\cN)$ is contained in the strong converse region $\Rc_\x{sc}$ of $\cN_{A'\to BC}$.
}

\begin{proof}
    Actually $C_g(\cN)$ is given by $C_g(\cN)=\log\min\{m_1m_2:g(\cN,m_1,m_2)\le 1\}$. Notice that $g$ is a strictly decreasing function of $m_1m_2$ in the sense that $g(\cN,m_1,m_2)>g(\cN,m_1',m_2')$ for $m_1m_2<m_1'm_2'$. Suppose $\log m_1+\log m_2>C_g(\cN)$, then $g(\cN,m_1,m_2)<1$, and $g(\cN,m_1,m_2)^n\to 0$ as $n\to\infty$. It follows that $f_\x{ua}(\cN^{\ot n},m_1^n,m_2^n)\le f_\x{NSPPT} (\cN^{\ot n},m_1^n,m_2^n)\le g(\cN^{\ot n},m_1^n,m_2^n) \le g(\cN,m_1,m_2)^n\to0$. By definition, $(\log m_1,\log m_2)$ is a strong converse rate pair.
\end{proof}

The authors of Ref. \cite{wang2017semidefinite} give a strong converse bound for the point-to-point classical communication, that is, $C(\cN)\le C_\bt(\cN)$, where $C(\cN)$ is asymptotic unassisted classical capacity of $\cN$. When $\cN$ is viewed as a point-to-point channel from $A'$ to $BC$, $C_\bt(\cN)$ is given by
\Als{
    C_\bt(\cN)=\log \x{min } & \Tr(S_{BC}) \\
    \x{s.t. } & \1_{A'} S_{BC} \ge R_{A'BC}\PT{BC} \ge -\1_{A'}S_{BC} \\
    & R_{A'BC}\ge J_\cN\PT{BC} \ge -R_{A'BC}.
}

\Prop{\label{prop-compare}
    For any quantum broadcast channel $\cN_{A'\to BC}$,
    \Eq{\label{compare-ineq}
        C_g(\cN)\le C_\bt(\cN).
        }
    In particular, this inequality can be strict for some channels.
}
\begin{proof}
    Suppose $(S^\star,R^\star)$ is an optimal solution to $C_\bt(\cN)$, it suffices to check that $\Tr S^\star$ is achievable by (\ref{sdp-fidelity-conv-dual-thm}). Choose $V=Y=Z=(R^\star)\PT{A'}$ and $Q=(S^\star)\T$. Since $\1_{A'} S_{BC}^\star \ge (R^\star)\PT{BC}\ge -\1_{A'} S_{BC}^\star$, one has $\1_{A'} (S_{BC}^\star)\T \ge (R^\star)\PT{A'}\ge -\1_{A'} (S_{BC}^\star)\T$, hence $\1_{A'} Q_{BC}\ge V_{A'BC} \ge -\1_{A'} Q_{BC}$. The second and third constraints in (\ref{sdp-fidelity-conv-dual-thm}) become trivial. It follows from $R_{A'BC}^\star \ge J_\cN\PT{BC} \ge -R_{A'BC}^\star$ that $Z_{A'BC}\PT{A'} \ge J_{\cN}\PT{BC} \ge -Z_{A'BC}\PT{A'}$.
    
% \begin{figure}[h]
% \centering
% \includegraphics[width=0.4\textwidth]{compare_N_r1} %.eps
% \caption{The broadcast capacity bound $C_g$ is strictly smaller than the point-to-point capacity bound $C_\bt$ for channels $\cN_r$.}
% \label{figure-compare-Nr}
% \end{figure}

\begin{figure}[h]
\centering
\begin{tikzpicture}[scale=0.84]
\begin{axis}[xmin=0,xmax=1, ymin=1.56,ymax=2,
            xlabel=Parameter $r$, ylabel=Capacity bound,
            legend style={at={(0.02,0.98)},anchor=north west}]
    \addplot [mark=none, black, very thick, dashed] table {broadcast_Nr_old.txt};
    \addplot [mark=none, black, very thick] table {broadcast_Nr_new.txt};
    \legend{$C_\bt(\cN_r)$, $C_g(\cN_r)$}
\end{axis}
\end{tikzpicture}
\caption{The broadcast capacity bound $C_g$ is strictly smaller than the point-to-point capacity bound $C_\bt$ for channels $\cN_r$.}
\label{figure-compare-Nr}
\end{figure}
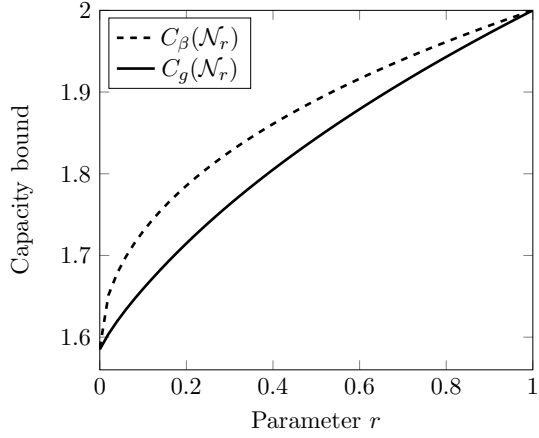

% \begin{figure}[h]
% \centering
% \includegraphics[width=0.45\textwidth]{compare-random} %.eps
% \caption{Plot of left-hand side of (\ref{compare-ineq}) vs. right-hand side for 1000 randomly generated channels $\cN$. We see the inequality (\ref{compare-ineq}) is strict for almost all points.}
% \label{figure-compare-random}
% \end{figure}

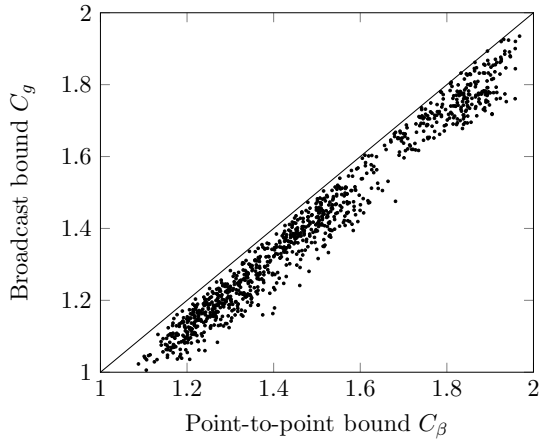
\begin{figure}[h]
\centering
\begin{tikzpicture}[scale=0.84]
\tikzset{mark options={mark size=0.6, line width=0pt}}
\begin{axis}[xmin=1,xmax=2, ymin=1,ymax=2,
            xlabel=Point-to-point bound $C_\bt$, ylabel=Broadcast bound $C_g$,
            mark=*]
    \addplot [mark=none, black] {x};
    \addplot [only marks,black] table {broadcast_random.txt};
\end{axis}
\end{tikzpicture}
\caption{Plot of left-hand side of (\ref{compare-ineq}) vs. right-hand side for 1000 randomly generated channels $\cN$. We see the inequality (\ref{compare-ineq}) is strict for almost all points.}
\label{figure-compare-random}
\end{figure}

    In some cases, our bound $C_g$ is strictly smaller than the bound $C_\bt$. For example, consider $\cN_r(\rho)=E_0\rho E_0^\dagger+E_1\rho E_1^\dagger$ with $E_0=\proj 0+\proj 1+\proj 2 + \sqrt r \proj 3$ and $E_1=\sqrt{1-r} \ketb 0 3$ ($0\leq r\leq 1$). Fig. \ref{figure-compare-Nr} shows that $C_g(\cN_r)<C_\bt(\cN_r)$ for this class of channels.
\end{proof}

We now randomly choose quantum broadcast channel $\cN_{A'\to BC}$ with $d_{A'}=4,d_B=d_C=2$, and compare our converse bound $C_g$ with the point-to-point bound in $C_\bt$, in Fig. \ref{figure-compare-random}. The numerical results suggest that $C_g<C_\bt$ holds for generic channel $\cN$.

\subsection{Converse bound on one-shot communication capacity based on hypothesis testing}
%\XW{Maybe get some insights from Th 5 of A. Anshu, R. Jain, and N. A. Warsi, arXiv:1702.01940.}
In this subsection, following the work \cite{wang2017converse}, we derive a converse bound on one-shot communication rate based on quantum hypothesis testing, and then obtain a converse bound on the asymptotic unassisted classical capacity. %A sub-channel is a completely positive map which is trace non-increasing.

Define
\Al{
\cQ_{BC}:=& \{\cF \in\x{CP}(A'\to BC): \exists \sg \in\cS(BC) \x{ s.t. }\cF(\rho_{A'})\le \sg ,\forall \rho \in\cS(A') \}, \\
\cQ_B:=& \{\cF \in\x{CP}(A'\to BC): \exists \sg \in\cS(B) \x{ s.t. }\Tr_C(\cF(\rho_{A'}))\le \sg ,\forall  \rho \in\cS(A') \}, \\
\cQ_C:=& \{\cF \in\x{CP}(A'\to BC): \exists \sg \in\cS(C) \x{ s.t. }\Tr_B(\cF(\rho_{A'}))\le \sg,\forall  \rho \in\cS(A') \},
}
where $\x{CP}(A'\to BC)$ denotes the set of completely positive linear maps from $A'$ to $BC$, and $\cS$ denotes the set of density operators.

The quantum hypothesis testing divergence \cite{wang2012one} is defined as
\Eq{
    D_\x{h}^\epsl(\rho\|\sg):=-\log\min\{\Tr(\sg T):\Tr(\rho T)\ge 1-\epsl,0\le T\le \1\}.
}

\Thm{
    For a quantum broadcast channel $\cN_{A'\to BC}$, if $(R_1,R_2)$ is in the unassisted one-shot $\epsl$-error classical capacity region, then
    \Al{
    R_1 &\le \min_{\cF\in\cQ_{B}}\max_{\phi}D_\x{h}^\epsl(\cN_{A'\to BC}(\phi_{\tilde AA'})\|\cF_{A'\to BC}(\phi_{\tilde AA'})), \label{pf-th10-1} \\
    R_2 &\le \min_{\cF\in\cQ_{C}}\max_{\phi}D_\x{h}^\epsl(\cN_{A'\to BC}(\phi_{\tilde AA'})\|\cF_{A'\to BC}(\phi_{\tilde AA'})), \label{pf-th10-2} \\
    R_1+R_2 &\le \min_{\cF\in\cQ_{BC}}\max_{\phi}D_\x{h}^\epsl(\cN_{A'\to BC}(\phi_{\tilde AA'})\|\cF_{A'\to BC}(\phi_{\tilde AA'})), \label{pf-th10-3}
    }
    where the maximum is over all pure state $\phi_{\tilde AA'}$, and $\tilde A$ and $A'$ are isomorphic systems.
}

% $R_i=\log m_i$

\begin{proof}
By hypothesis we have $f_\x{ua}(\cN,m_1,m_2)\ge 1-\epsl$ for $m_1=2^{R_1}$ and $m_2=2^{R_2}$.

% It suffices to show that if  then the following hold,
% \Al{
%     \log m_1 &\le \min_{\cF\in\cQ_{B}}\max_{\phi}D_\x{h}^\epsl(\cN(\phi)\|\cF(\phi)), \label{pf-th10-1} \\
%     \log m_2 &\le \min_{\cF\in\cQ_{C}}\max_{\phi}D_\x{h}^\epsl(\cN(\phi)\|\cF(\phi)), \label{pf-th10-2} \\
%     \log m_1+ \log m_2 &\le \min_{\cF\in\cQ_{BC}}\max_{\phi}D_\x{h}^\epsl(\cN(\phi)\|\cF(\phi)). \label{pf-th10-3}
% }

In order to show Eq. (\ref{pf-th10-1}), we need to show that for all $\cF\in\cQ_B$, there exists $\phi_{\tilde AA'}$ and $0\le T_{\tilde ABC}\le \1$, such that $\Tr(\cN_{A'\to BC}(\phi_{\tilde AA'})T_{\tilde ABC})\ge 1-\epsl$ and $\Tr(\cF_{A'\to BC}(\phi_{\tilde AA'})T_{\tilde ABC})\le \f{1}{m_1}$.

Since $f_\x{ua}(\cN,m_1,m_2)\ge 1-\epsl$, there is a choice of coding scheme such that $\f{1}{m_1m_2}\sum_{i,j=1}^{m_1,m_2} \Tr(\cN(\rho_{ij,A'})(E_{i,B}\ot F_{j,C})) \ge 1-\epsl$. Here $\{\rho_{ij,A'}\}_{i,j=1}^{m_1,m_2}$ is input state set of $\cN$ as encoding scheme, and two POVMs $\{E_{i,B}\}_{i=1}^{m_1}$, $\{F_{j,C}\}_{j=1}^{m_2}$ serve as decoding scheme. Notice that
\Als{
1-\epsl &\le \f{1}{m_1m_2}\sum_{i,j=1}^{m_1,m_2} \Tr(\cN_{A'\to BC}(\rho_{ij,A'})(E_{i,B}\ot F_{j,C})) \\
&= \f{1}{m_1m_2}\sum_{i,j=1}^{m_1,m_2} \Tr \big(J_{\cN,A'BC}(\rho_{ij,A'}\T\ot E_{i,B}\ot F_{j,C}) \big)  \\
&=  \f{1}{m_1m_2}\sum_{i,j=1}^{m_1,m_2} \Tr \Big( \cN_{A'\to BC}(\phi_{\tilde AA'})(\hat\rho_{\tilde A}\T)^{-1/2}(\rho_{ij,\tilde A}\T\ot E_{i,B}\ot F_{j,C})(\hat\rho_{\tilde A}\T)^{-1/2} \Big),
}
where $\hat\rho_{\tilde A}=\f{1}{m_1m_2}\sum_{i,j}\rho_{ij,\tilde A}$ with $\rho_{ij,\tilde A}=\id_{A'\to\tilde A} (\rho_{ij,A'})$, and $\phi_{\tilde AA'}=\ketb{\phi}{\phi}$ with $\ket{\phi}=\vc \big( \1_{\tilde A\to A'}\hat\rho_{\tilde A}^{1/2} \big)$. Take the tester $T=(\hat\rho_{\tilde A}\T)^{-1/2}  \sum_{ij}\f{1}{m_1m_2}(\rho_{ij,\tilde A}\T\ot E_{i,B}\ot F_{j,C}) (\hat\rho_{\tilde A}\T)^{-1/2}$, and it thus holds that $0\le T\le \1$ and $\Tr(\cN_{A'\to BC}(\phi_{\tilde AA'})T)\ge 1-\epsl$. Now the Eq. (\ref{pf-th10-1}) follows.

% DETAILS: $J_{\cN,A'BC}=(\id_{\tilde A\to A'}\ot\cN_{A'\to BC})\ketb{\psi_{\tilde AA'}}{\psi_{\tilde AA'}}$ where $\psi$ is the unnormalised isotropic maximally entangled state. It can be also written as $\ket{\psi_{\tilde AA'}}=\sg_{\tilde A}^{\s{T},-1/2} \vc (\1_{\tilde A\to A'} \sg_{\tilde A}^{1/2})$ .

For any $\cF\in\cQ_B$,
\Als{
\Tr(\cF_{A'\to BC}(\phi_{\tilde AA'})T_{\tilde ABC}) & = \f{1}{m_1m_2}\sum_{i,j=1}^{m_1,m_2} \Tr(\cF(\rho_{ij})(E_i\ot F_j)) \\
& \le \f{1}{m_1m_2}\sum_{i,j=1}^{m_1,m_2} \Tr(\cF(\rho_{ij})(E_i\ot \1)) \\
& = \f{1}{m_1m_2}\sum_{i,j=1}^{m_1,m_2} \Tr(\Tr_C(\cF(\rho_{ij}))E_i) \\
& \le \f{1}{m_1m_2}\sum_{i,j=1}^{m_1,m_2} \Tr(\sg_BE_i) = \f{1}{m_1}.
}

Similarly, for $\cF\in\cQ_C$, $\Tr(\cF(\phi)T) \le\f{1}{m_2}$, and for $\cF\in\cQ_{BC}$, $\Tr(\cF(\phi)T) \le\f{1}{m_1m_2}$. Therefore Eqs. (\ref{pf-th10-1}), (\ref{pf-th10-2}) and (\ref{pf-th10-3}) hold for any $m_1,m_2$ satisfying $f_\x{ua}(\cN,m_1,m_2)\ge 1-\epsl$.
\end{proof}

\section{Classical communication over quantum multi-access channel}\label{sec:mac}

\begin{figure}[htbp]
\centering
\begin{tikzpicture}
    \def\step{0.2cm}  \def\xo{0.06cm}
    % \draw[help lines,dashed,step=\step,xshift=0cm, yshift=0cm] (0,0) grid (4.6,2.8);
    \draw[black,fill=black!33,opacity=0.4] (2.2,0) rectangle (3.4,0.8) node[midway,black,opacity=1] {$\cN$};
    \draw (0.2,1.6) rectangle (1,2.6) node[midway] {$\cE_1$};
    \draw (1.2,1.6) rectangle (2,2.6) node[midway] {$\cE_2$};
    % \draw (2.6,1.6) rectangle (3.4,2.6) node[midway] {$\cD_1$};
    \draw (3.6,1.6) rectangle (4.4,2.6) node[midway] {$\cD$};
    \draw[thick,->] (0.6,3.4) -- node[right,xshift=-\xo,yshift=\step] {$A $} (0.6,2.6);
    \draw[thick,->] (1.6,3.4) -- node[right,xshift=-\xo,yshift=\step] {$B$} (1.6,2.6);
    \draw[thick,<-] (4  ,3.4) -- node[right,xshift=-\xo,yshift=\step] {$C'$} (4,2.6);
    \draw[thick,->] (0.6,1.6) -- node[right,xshift=-\xo,yshift=\xo] {$A'$} (0.6,0.33) -- (2.2,0.33);
    \draw[thick,->] (1.6,1.6) -- node[right,xshift=-\xo] {$B'$} (1.6,0.47) -- (2.2,0.47);
    \draw[thick,->] (3.4,0.4) -- (4,0.4) -- node[right,xshift=-\xo,yshift=0.05cm] {$C$} (4,1.6);
    \draw[black,fill=black!33,opacity=0.4] (0,1.3) rectangle (4.6,2.8) node[midway,black,opacity=1,xshift=0.5cm] {\large $\cX$};
    \draw[thick,dashed,black!70] (-0.2,-0.1) node[xshift=0.24cm,yshift=0.17cm] {$\cM$} rectangle (4.8,2.9);
\end{tikzpicture}
\caption{Classical communication over quantum multi-access channel $\cN$ assisted by general code $\cX$.}
\label{fig:2-to-1}
\end{figure}
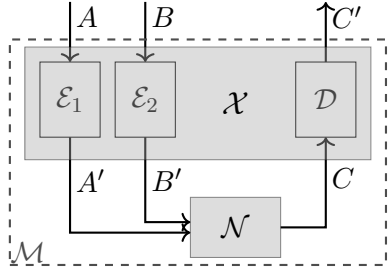

We have explored the classical communication over the quantum broadcast channel. Using similar approaches we can study the quantum multi-access channel; see Fig. \ref{fig:2-to-1}. We now derive a strong converse rate region of quantum multi-access channel $\cN_{A'B'\to C}$.
 % for the model. The Definitions \label{def1,def2,def3} can apply to the quantum multi-access channel with slight change.

Similar to Def. \ref{def1}, the success probability of $\cN_{A'B'\to C}$ to transmit messages of size $(m_1,m_2)$ assisted by code $\cX_{ABC\to A'B'C'}$ is defined as
    \Eq{
        p_s(\cN,\cX,m_1,m_2) = \f{1}{m_1m_2}\sum_{i,j=1}^{m_1,m_2} \Tr(\cM(\ketb{ij}{ij}_{AB})\ketb{ij}{ij}_{C'}),
    }
where $\cM_{AB\to C'}=\cN\circ\cX$. The $\Om$-assisted optimal success probability is similarly defined as $f_{\Om}(\cN,m_1,m_2)=\max_{\cX\in\Om} p_s(\cN,\cX,m_1,m_2)$. $(R_1,R_2)$ is called a strong converse rate pair for $\cN$ if
\Eq{
\lim_{n\to\infty} f_\x{ua} (\cN^{\ot n},2^{nR_1},2^{nR_2})=0. }

We give the following characterization with proof put in Appendix \ref{app-a}.

\Thm{\label{thm-fidelity-qmac}
    The optimal success probability of quantum multi-access channel $\cN_{A'B'\to C}$ to transmit messages of size $(m_1,m_2)$ assisted by non-signalling and PPT codes is given by
    \Als{\label{fidelity-qmac}
    f_\x{NSPPT}(\cN,m_1,m_2) = \x{max } & \Tr(J_{\cN,A'B'C}\T E_{1,A'B'C}) \\
    \x{s.t. } & E_i,E_i\PT{A'},E_i\PT{B'},E_i\PT{C}\ge 0, \\
    & \Tr L_{A'B'}=d_C, \x{ (TP)} \\
    & L_{A'B'C}=\f{\1_C}{d_C}L_{A'B'}, \ (C\not\to AB) \\
    & E_{i,C}=\f{\1_C}{m_1m_2}, \ (AB\not\to C) \\
    & E_{1,A'C}=E_{2,A'C}, E_{3,A'C}=E_{4,A'C}, \ (B\not\to AC) \\
    & E_{1,B'C}=E_{3,B'C}, E_{2,B'C}=E_{4,B'C}, \ (A\not\to BC)
    }
    where $L_{A'B'C}:=E_{1,A'B'C}+(m_2-1)E_{2,A'B'C}+(m_1-1)E_{3,A'B'C}+(m_1-1)(m_2-1)E_{4,A'B'C}$.
}

Similar to Prop. \ref{prop:g(N)}, we have a sub-multiplicative converse bound for quantum multi-access channels as follows. See Appendix \ref{app-a} for the proof.

\Prop{\label{prop:g(N)-qmac}
    For any quantum multi-access channel $\cN_{A'B'\to C}$ and given $m_1,m_2$,
    \Eq{
    f_\x{NSPPT}(\cN,m_1,m_2) \le h(\cN,m_1,m_2),
    }
    where
    \Als{
        \label{sdp-fidelity-conv-dual2-qmac}
        h(\cN,m_1,m_2) := \x{min } & \Tr Q_C \\
        \x{s.t. } & m_1m_2\1_{A'B'}Q_C \ge V_{A'B'C} \ge -m_1m_2\1_{A'B'}Q_C, \\
        & V_{A'B'C}\PT{A'} \ge Y_{A'B'C}\PT{A'} \ge -V_{A'B'C}\PT{A'}, \\
        & Y_{A'B'C}\PT{B'} \ge Z_{A'B'C}\PT{B'} \ge -Y_{A'B'C}\PT{B'}, \\
        & Z_{A'B'C}\PT{C} \ge J_{\cN}\PT{A'B'} \ge -Z_{A'B'C}\PT{C}.
    }
    
    Furthermore, $h$ is submultiplicative in the sense that $h(\cN\ot\cN',m_1m_1',m_2m_2') \le h(\cN,m_1,m_2) h(\cN',m_1',m_2')$. Consequently, $f_\x{NSPPT}(\cN^{\ot n},m_1^n,m_2^n)\le h(\cN,m_1,m_2)^n$.
}

Similar to Theorem \ref{thm:converse}, we have

\Thm{\label{thm:converse-qmac}
    For a quantum multi-access channel $\cN_{A'B'\to C}$, if $R_1+R_2 > C_h(\cN)$, then $(R_1,R_2)$ is a strong converse rate pair. Here
    \Als{\label{sdp-C-h}
        C_h(\cN) := \log \x{min } & \Tr Q_C \\
        \x{s.t. } & \1_{A'B'}Q_C \ge V_{A'B'C} \ge -\1_{A'B'}Q_C, \\
        & V_{A'B'C}\PT{A'} \ge Y_{A'B'C}\PT{A'} \ge -V_{A'B'C}\PT{A'}, \\
        & Y_{A'B'C}\PT{B'} \ge Z_{A'B'C}\PT{B'} \ge -Y_{A'B'C}\PT{B'}, \\
        & Z_{A'B'C}\PT{C} \ge J_{\cN}\PT{A'B'} \ge -Z_{A'B'C}\PT{C}.
    }
}

\begin{proof}
    Indeed $C_h(\cN)$ is given by $C_h(\cN)=\log\min\{m_1m_2:h(\cN,m_1,m_2)\le 1\}$. Notice that $h$ is a strictly decreasing function of $m_1m_2$. Suppose $\log m_1+\log m_2>C_h(\cN)$, then $h(\cN,m_1,m_2)<1$, and $h(\cN,m_1,m_2)^n\to 0$ as $n\to\infty$. It follows that
    \Als{
        f_\x{ua}(\cN^{\ot n},m_1^n,m_2^n) &\le f_\x{NSPPT} (\cN^{\ot n},m_1^n,m_2^n) \\
        &\le h(\cN^{\ot n},m_1^n,m_2^n) \\
        &\le h(\cN,m_1,m_2)^n\to 0.
        }
    Then $(\log m_1,\log m_2)$, by definition, is a strong converse rate pair of multi-access channel $\cN$.
\end{proof}

We find that $C_h$ is a better bound than $C_\bt$ for quantum multi-access channels. In this case $C_\bt$ is written as
\Als{\label{C-beta-qmac}
    C_\bt(\cN)=\log \x{min } & \Tr(S_C) \\
    \x{s.t. } & \1_{A'B'} S_C \ge R_{A'B'C}\PT{C} \ge -\1_{A'B'} S_C \\
    & R_{A'B'C}\ge J_\cN\PT{C} \ge -R_{A'B'C}.
}

\Prop{
    For any quantum multi-access channel $\cN_{AB\to C}$,
    \Eq{\label{compare-ineq-qmac}
        C_h(\cN)\le C_\bt(\cN).
        }
    In particular, this inequality can be strict for some channels.
}

\begin{figure}[h]
\centering
\begin{tikzpicture}[scale=0.84]
\begin{axis}[xmin=0,xmax=1, ymin=1.56,ymax=2,
            xlabel=Parameter $r$, ylabel=Capacity bound,
            legend style={at={(0.02,0.98)},anchor=north west}]
    \addplot [mark=none, black, very thick, dashed] table {qmac_Nr_old.txt};
    \addplot [mark=none, black, very thick] table {qmac_Nr_new.txt};
    \legend{$C_\bt(\cN_r)$, $C_h(\cN_r)$}
\end{axis}
\end{tikzpicture}
\caption{The multi-access capacity bound $C_h$ is strictly smaller than the point-to-point capacity bound $C_\bt$ for channels $\cN_r$.}
\label{figure-qmac-compare-Nr}
\end{figure}
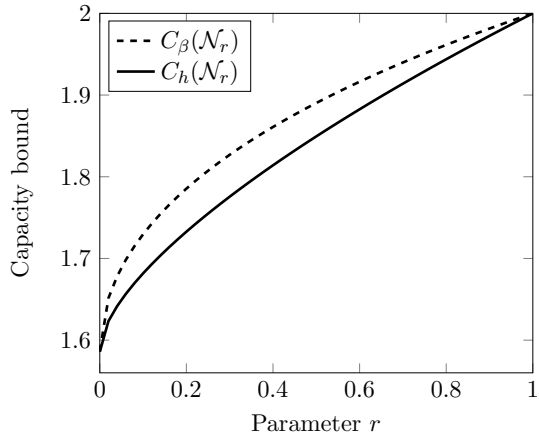

\begin{proof}
    Similar to the proof of Proposition \ref{prop-compare}, we take $V=Y=Z=R\PT{A'B'}$ and $Q=S\T$, provided $(S_C^\star,R_{A'B'C}^\star)$ is an optimal solution to SDP (\ref{C-beta-qmac}). So the optimal value $\Tr(S_C^\star)$ can be achieved by the SDP (\ref{sdp-C-h}) of $C_h$. Thus the inequality holds.

    Considering the channel $\cN_r$ defined in Prop. \ref{prop-compare}, Fig. \ref{figure-qmac-compare-Nr} shows that $C_h(\cN_r) < C_\bt(\cN_r)$ holds for this class of channels. The figure looks quite similar to fig \ref{figure-compare-Nr}, but the values are indeed different. In fact numerical results suggest that this inequality is also strict for generic channels.
\end{proof}

% \begin{figure}[htbp]
% \centering
% \includegraphics[width=0.4\textwidth]{qmac_compare_N_r} %.eps
% \caption{The bound $C_h$ is strictly smaller than the point-to-point capacity bound $C_\bt$ for channels $\cN_r$.}
% \label{figure-qmac-compare-Nr}
% \end{figure}

\section{Discussions}\label{sec:last}

In summary, we have characterized the one-shot optimal average success probability of NS and PPT codes in classical communication over a given quantum network channel, in the form of SDP. By invoking the property of the SDPs, we have established strong converse rates for general quantum broadcast and multi-access channels. We also have obtained converse bounds for the one-shot  classical communication over network channels.

Although we did not deal with the channels with more than two senders or receivers, it can be expected to be a simple extension with more technical involvement. The approach used in our work can apply to the study of the quantum capacity of quantum channels and multi-partite entanglement distillation. It would be interesting to study further the property of the hypothesis testing divergence between a channel and a certain class of channels, which may help to explore more properties of the broadcast channel capacity region.

\section*{Acknowledgement}
This work was partly supported by the Australian Research Council under Grant Nos. DP120103776 and FT120100449.

\bibliography{bib} 

\appendix
\section{Proofs of Theorem \ref{thm-fidelity-qmac} and and Proposition \ref{prop:g(N)-qmac}}\label{app-a}
\begin{proof}[Proof of Theorem \ref{thm-fidelity-qmac}]
The success probability of code $\cX_{ABC\to A'B'C'}$ to transmit messages of size $(m_1,m_2)$ over channel $\cN_{A'B'\to C}$ is
\Al{
    p_s(\cN,\cX,m_1,m_2) &= \f{1}{m_1m_2} \sum_{i,j=1}^{m_1,m_2} \Tr(\ketb{ij}{ij}(\cN\circ\cX)(\ketb{ij}{ij}) ) \\
    % &= \f{1}{m_1m_2} \Tr(J_{\cM,ABC'}\T(D_{AC'_1}\ot D_{BC'_2})) \\
    &= \f{1}{m_1m_2} \Tr(J_{\cN,ABC'}\T  X_{ABCA'B'C'}(D_{1,AC'_1}\ot D_{2,BC'_2})) \\
    &= \f{1}{m_1m_2} \Tr(J_{\cN,ABC'}\T \tilde X_{ABCA'B'C'}(D_{1,AC'_1}\ot D_{2,BC'_2})),
}
where $D_{1,AC'_1}=\sum_{k=1}^{m_1}\ketb{kk}{kk}$ and $D_{2,BC'_2}=\sum_{k=1}^{m_2}\ketb{kk}{kk}$ and $\tilde X=\Tc_{AC'_1}\Tc_{BC'_2}(X)$. The optimal success probability of $\cN$ to transmit messages of size $(m_1,m_2)$ assisted by $\Om$-class codes is defined as
\Eq{
    f_\Om(\cN,m_1,m_2) = \max_{\cX\in\Om} p_s(\cN,\cX,m_1,m_2).
}

Since the registers $A,B,C'$ are classical, we have
\Als{
\tilde X= & D_{1,AC'_1}D_{2,BC'_2}E_{1,A'B'C}+D_{1,AC'_1}(\1-D_{2,BC'_2})E_{2,A'B'C} \\
 & +(\1-D_{1,AC'_1}) D_{2,BC'_2}E_{3,A'B'C}+(\1-D_{1,AC'_1}) (\1-D_{2,BC'_2})E_{4,A'B'C}.
}

Now let us consider the  NSPPT codes. $\tilde \cX$ is CP and PPT iff $E_i\ge0,E_i\PT{A'}\ge0,E_i\PT{B'}\ge0,E_i\PT{C}\ge0$ for $i=1,2,3,4$. Denoting $L_{A'B'C}=E_{1,A'B'C}+(m_2-1)E_{2,A'B'C}+(m_1-1)E_{3,A'B'C}+(m_1-1)(m_2-1)E_{4,A'B'C}$, the CP constraint is $L_C=\1_C$.

The NS constraint $C\not\to AB$ is $\tilde X_{ABCA'B'}=\f{\1_C}{d_C}\tilde X_{ABA'B'}$, equivalent to $L_{A'B'C}=\f{\1_C}{d_C}L_{A'B'}$. Thus the TP condition becomes $\Tr L_{A'B'}=d_C$. The NS constraint $AB\not\to C$ is equivalent to $E_{i,C}=\f{\1_C}{m_1m_2}$. The constraints $AC\not\to B$ and $BC\not\to A$ are implicitly implied. The constraint $B\not\to AC$ is $\tilde X_{ABCA'C'}=\f{\1_B}{d_B} \tilde X_{ACA'C'}$, which is equivalent to $E_{1,A'C}=E_{2,A'C}$ and $E_{3,A'C}=E_{4,A'C}$. Similarly the NS constraint $A\not\to BC$ is $E_{1,B'C}=E_{3,B'C}$ and $E_{2,B'C}=E_{4,B'C}$.

Putting together the above constraints, we obtain the SDP in Theorem \ref{thm-fidelity-qmac}.
\end{proof}

\begin{proof}[Proof of Proposition \ref{prop:g(N)-qmac}]
        It is easy to see
    \Als{
        f_\x{NSPPT}(\cN,m_1,m_2) \le \x{max } & \Tr(J_{\cN,A'B'C}\T E_{1,A'B'C}) \\
                \x{s.t. } & E_1,E_1\PT{A'},E_1\PT{B'},E_1\PT{C}\ge 0, \\
                & E_{1,C}=\f{\1_C}{m_1m_2}.
        }
    The dual of the right-hand-side SDP is
    \Als{\label{qmac-sdp1}
        \x{min } & \Tr Q_C \\
        \x{s.t. } & J_\cN\T+P_1\PT{A}+P_2\PT{B}+P_3\PT{C}\le m_1m_2\1_{AB}Q_C \\
        & P_i\ge 0.
    }
    Introducing $V:=J_\cN\T+P_1\PT{A}+P_2\PT{B}+P_3\PT{C}$, $Y:=J_\cN\T+P_2\PT{B}+P_3\PT{C}$ and $Z:=J_\cN\T+P_3\PT{C}$, we have the SDP (\ref{qmac-sdp1}) is equivalent to 
    \Als{
        \x{min } & \Tr Q_C \\
        \x{s.t. } & m_1m_2\1_{AB}Q_C \ge V, \\
        & V\PT{A} \ge Y\PT{A}, \\
        & Y\PT{B} \ge Z\PT{B}, \\
        & Z\PT{C} \ge J_\cN\PT{AB}.
    }
    By adding new constraints, the above SDP is no larger than $h(\cN,m_1,m_2)$. It can be readily verified that $h$ is sub-multiplicative.
\end{proof}

\end{document}